\documentclass[letterpaper, 9 pt, conference]{ieeeconf}

\IEEEoverridecommandlockouts
\usepackage{color}
\usepackage{enumerate,graphicx,epstopdf}
\usepackage{amsmath,amssymb,bm}
\usepackage{cite}
\usepackage{mathtools}
\usepackage{array}

\newtheorem{rmk}{Remark}
\newtheorem{prb}{Problem}
\newtheorem{dfn}{Definition}

\newtheorem{prp}{Proposition}

\newcommand{\real}{\mathbb{R}}
\newcommand{\sqr}{$\hfill\square$}

\title{Explicit Reference Governor for the\\ Constrained Control of Time-Delayed Linear Systems
\thanks{This research is supported by the National Science Foundation, award number CMMI 1562209, and the FNRS MIS research grant number F.4526.17 F.}
\thanks{M.M. Nicotra and I.V. Kolmanovsky are with the Department of Aerospace Engineering at the University of Michigan, Ann Arbor (MI), United States. Email: \texttt{\{mnicotra,ilya\}@umich.edu}.}
\thanks{E. Garone and T. Nguyen are with the Service d'Automatique et d'Analyse des Syst\`emes at the Universit\'e Libre de Bruxelles, Brussels, Belgium. Email: \texttt{\{egarone,tanguyen\}@ulb.ac.be}.}
}
\author{Marco M. Nicotra, Tam Nguyen, Emanuele Garone, Ilya V. Kolmanovsky}

\begin{document}
\maketitle

\begin{abstract}
This paper introduces an explicit reference governor approach for controlling time delay linear systems subject to state and input constraints. The proposed framework relies on suitable invariant sets that can be built using both Lyapunov-Razumikhin and Lyapunov-Krasovskii arguments. The proposed method is validated both numerically and experimentally using several alternative formulations.
\end{abstract}

\section{Introduction}
Time delay systems are a class of dynamic systems in which the reaction to an exogenous input is not instantaneous. This behavior is often encountered due to the presence of propagation and transport phenomena affecting the sensors, actuators, or communication channels. The stability and stabilization of time delay systems has been studied extensively in the literature, see e.g. \cite{TDS1,TDS2,TDS3,FridmanBook}. 

In the presence of input saturation constraints, \cite{ActSat,ActSat2,ActSat3} propose control design methods that ensure a satisfactory basin of attraction. To address the presence of state constraints, \cite{D-Invariance,InvSet} illustrate how to construct polyhedral forward-invariant sets for a given pre-stabilized system. Although these \textit{a posteriori} analyses can be useful for tuning the feedback gain and increasing the set of admissible initial conditions, the resulting control laws do not actively guarantee constraint satisfaction.

For what concerns control methods that actively enforce constraints, the typical approach consists in formulating an optimal control problem and solving it in real-time. Model Predictive Control (MPC) schemes for time delay system subject to input constraints have been proposed in \cite{MPC1,MPC2,MPC3,MPC4}. Additionally, Reference Governor (RG) schemes have been proposed in \cite{Bemporad1998,DiCairanoKalabic2015,LiKalabic2014,CasavolaMosca2006,CasavolaPapini2006}. The main difference between MPC and RG is that the former generates the control input based on the predicted system trajectories, whereas the latter generates a reference for the pre-stabilized system and is typically less computationally expensive due to its formulation \cite{RG_Survey}. The main drawback of these approaches is that the computational cost of solving the optimal control problem is non negligible and may sometimes be prohibitive for embedded hardware.

This paper introduces an Explicit Reference Governor (ERG) scheme that actively enforces state and input constraints without the need to solve an optimization problem. As detailed in \cite{ERGbasic}, the main idea behind the ERG approach is that it is possible to ensure constraint satisfaction by pre-stabilizing the system and introducing an \textit{add-on} unit that suitably manipulates the derivative of the applied reference. Since the ERG is based on set invariance principles, this paper investigates the behavior that can be obtained using different Lyapunov-Razumikhin and Lyapunov-Krasovskii level-sets. The proposed scheme is validated both numerically and experimentally.

\section{Problem Statement}

In this paper we consider linear systems with input delay of the form,
\begin{equation}\label{eq:System}
\begin{cases}
\dot{x}(t)=Ax(t)+Bu(t-\tau),\\
y(t) = Cx(t)+Du(t-\tau),
\end{cases}
\end{equation}
where $x(t)\in\real^n$ is the state, $u(t)\in\real^m$ is the control input, $y(t)\in\real^p$ is the output, $\tau>0$ is a constant delay, and $(A,B,C,D)$ are suitably dimensioned matrices such that the system is stabilizable.  System \eqref{eq:System} is subject to linear inequality constraints on both the state and 
input vectors,
\begin{equation}\label{eq:OriginalConstraints}
h_{x,i}^T\,x(t)+h_{u,i}^Tu(t)+g_i\geq0,\quad i=1,\ldots,n_c.
\end{equation}
Given an output reference signal $r(t)\in\real^p$ and given suitable initial conditions, the objective of this paper is to design a closed-form control strategy such that
\begin{enumerate}[1.]
\item For any piece-wise continuous signal $r(t)$, not known in advance, constraints \eqref{eq:OriginalConstraints} are satisfied at all times;
\item If there exists $t_1$ such that $r(t)=r$ for $t \geq t_1$, and if the reference $r$ is consistent with the constraints \eqref{eq:OriginalConstraints}, then $\lim_{t\to\infty}y(t)=r$.
\end{enumerate}

\section{Control Strategy}
\begin{figure}
\center
\includegraphics[width=\columnwidth]{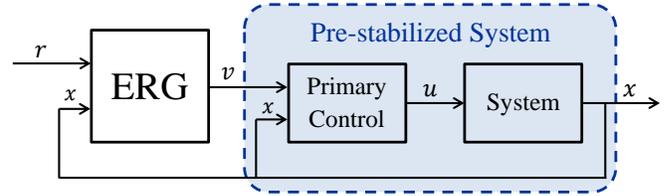}
\vspace{-0.5 cm}
\caption{General architecture used by the ERG framework: the system is pre-stabilized using a primary control loop and the ERG is used as an \textit{add-on} unit that provides constraint handling capabilities.\label{fig:ERG_Scheme}}
\end{figure}
Following the reference governor philosophy, this paper proposes a two-step approach for performing the constrained control of a time delay system. The first step consists in pre-stabilizing the system dynamics using a primary control loop. The second step then ensures constraint enforcement with the aid of an auxiliary control loop based on the ERG. Referring to Figure \ref{fig:ERG_Scheme}, this is achieved by introducing an auxiliary reference signal $v(t)\in\real^p$ to decouple the two control loops. The objective of the primary control loop is to solve the following control problem.

\smallskip
\begin{prb}\label{prb:PrimaryControl}
Consider a constant reference $v\in\real^p$, and let $\bar{x}_v$, $\bar{u}_v$ be a steady-state equilibrium satisfying
\begin{eqnarray}
A\bar{x}_v+B\bar{u}_v&\!\!\!=&\!\!\!0,\\
C\bar{x}_v+D\bar{u}_v&\!\!\!=&\!\!\!v.
\end{eqnarray}
The objective of the primary control loop is to design a control feedback such that the equilibrium point $\bar{x}_v$, $\bar{u}_v$ is Globally Asymptotically Stable (GAS).\sqr
\end{prb}
\smallskip

The objective of the auxiliary control loop is to augment the primary control law by providing constraint handling capabilities. This is achieved by solving the following control problem.

\smallskip
\begin{prb}
Given the pre-stabilized system obtained as the solution to Problem \ref{prb:PrimaryControl}, and given suitable initial conditions $\bm{x}_\tau(0)$, $v(0)$, construct an auxiliary reference signal $v(t)$ such that:
\begin{enumerate}[1.]
\item For any piecewise continuous reference signal $r(t)\in\real^p$, the trajectories of $x(t)$, $u(t)$ satisfy constraints \eqref{eq:OriginalConstraints}.
\item For any constant reference signal $r(t)=r\in\real^p$, that strictly satisfies constraints \eqref{eq:OriginalConstraints}, $v(t)$ asymptotically tends to $r$.\sqr
\end{enumerate}
\end{prb}
\smallskip

The following section addresses the design of the primary control loop using well-known results from the existing literature. Particular emphasis is given to the characterization of suitable Lyapunov level-sets for the resulting pre-stabilized system. This will be used in Section \ref{sec:AuxiliaryControl} which addresses the design of the auxiliary control loop. 

\section{Primary Control}\label{sec:PrimaryControl}
The stabilization of Linear Time-Delayed (LTD) systems not subject to constraints is the subject of an extensive literature and can be addressed using a variety of existing tools. Given a constant reference $v$ and a linear control law in the form 
\begin{equation}\label{eq:PrimaryControl}
u(t)=\bar{u}_v+K(x(t)-\bar{x}_v),
\end{equation}
the dynamics of the closed-loop system satisfy
\begin{equation}\label{eq:PrestabilizedSystem}
\dot{e}(t)=Ae(t)+BKe(t-\tau),
\end{equation}
where $e(t)=x(t)-\bar{x}_v$. Problem \ref{prb:PrimaryControl} can therefore be reduced to finding a control gain $K$ such that the origin of \eqref{eq:PrestabilizedSystem} is GAS. Techniques for the stability analysis of time-delayed systems are described in, e.g. \cite{Fridman2014} and references therein. For the reader's convenience, the following subsections provide a brief summary of some of these results and characterizes the associated invariance properties.

\subsection{Stability Conditions without Delay}
In the absence of delays, i.e. $\tau=0$, it is well-known that the equilibrium point $\bar{x}_v$ is GAS if and only if it is possible to compute a quadratic function 
\begin{equation}\label{eq:LyapFun}
W(e)=e^TP\,e,
\end{equation}
such that $P>0$ satisfies the Linear Matrix Inequality (LMI)
\begin{equation}\label{eq:LyapIneq}
(A+BK)^T\!P+P(A+BK)<0.
\end{equation}
Indeed, the satisfaction of \eqref{eq:LyapIneq} is a sufficient (and necessary) condition for ensuring the Lyapunov criterion $\dot{W}(e)\leq-\gamma(\|e\|)$, where $\gamma$ is a class $\mathcal{K}_\infty$ function.

\subsection{Delay-Independent Conditions}
In the presence of a time delay $\tau>0$, it is not possible to construct a quadratic function \eqref{eq:LyapFun} which is monotonically time-decreasing. However, if there exists $P>0$ and $q>0$ satisfying the LMI
\begin{equation}\label{eq:LMI_Razumikhin}
\begin{bmatrix}
A^TP+PA+qP & PBK\\
* & -qP
\end{bmatrix}<0,
\end{equation}
it can be proven, see e.g. \cite[Theorem 3.2]{FridmanBook}, that \eqref{eq:LyapFun} satisfies the Lyapunov-Razumikhin criterion
\[
\dot{W}(e(t))\leq-\gamma(\|e(t)\|),~\text{if }W(e(t\!+\!\theta))\leq \rho W(e(t)),~\forall \theta\in[-\tau,0],
\]
for some $\rho>1$. This implies that \eqref{eq:LyapFun} is a Razumikhin function and that \begin{equation}\label{eq:LyapRazumikhin}
V(\bm{e_\tau}(t))=\max_{\theta\in[-\tau,0]}\{e(t\!+\!\theta)^TP\,e(t\!+\!\theta)\},
\end{equation} 
where
\begin{equation}\label{eq:Def_eTau}
\bm{e_\tau}(t)=\{e(t+\theta)\},\qquad \theta\in[-\tau,0],
\end{equation}
is a Lyapunov-Krasovskii functional\footnote{The notation $V(\bm{e_\tau}(t))$ is used 
as $V$ is not simply a function of the current state $e(t)$, but rather is 
a functional that depends on the state trajectories $e(\theta)$ within the past interval $[t-\tau,t]$.} for the pre-stabilized system.

Unlike the case $\tau=0$, the LMI \eqref{eq:LMI_Razumikhin} is only a \emph{sufficient} condition for proving GAS. As a result, it may be possible to prove stability for a wider range of systems by using more complex functions. A possible example is the Lyapunov-Krasovkii functional
\begin{equation}\label{eq:LyapFunctional_Q}
V(\bm{e_\tau}(t))=e(t)^TP\,e(t)+\int_{t-\tau}^{t}\!\!\!\!e(s)^TQ\,e(s)ds,
\end{equation}
with $P>0$, $Q>0$ satisfying the LMI
\begin{equation}\label{eq:LMI_LKQ}
\begin{bmatrix}
A^TP+PA+Q & PBK\\
* & -Q
\end{bmatrix}<0.
\end{equation}
The idea behind equation \eqref{eq:LyapFunctional_Q} is that the fluctuations of the classical quadratic term are compensated by the integral term to ensure that $V(\bm{e}_\tau(t))$ is monotonically time-decreasing. The advantage of this method is that \eqref{eq:LMI_LKQ} is less restrictive with respect to the Razumikhin conditions \eqref{eq:LMI_Razumikhin} since \eqref{eq:LMI_LKQ} also admits the solution $Q=qP$.\smallskip

It is worth noting that both the LMIs \eqref{eq:LMI_Razumikhin} and \eqref{eq:LMI_LKQ} can only be satisfied if there exists $S>0$ such that $A^TS+SA<0$. This implies that both methods require the open-loop system \eqref{eq:System} to be GAS. Although fairly restrictive, 
an appealing aspect of these methods is that \eqref{eq:LMI_Razumikhin} and \eqref{eq:LMI_LKQ} do not depend on the actual value of $\tau$, and hence 
they yield GAS for arbitrarily large time delays.

\subsection{Delay-Dependent Conditions}
In many applications, it is common to observe that a closed-loop system may become unstable given an excessive time delay. As a result, it is reasonable to assume that, for many LTD systems, the stability conditions should also depend on the value of $\tau$. In this setting, a well-known (albeit simple) approach for proving GAS is the use of the Lyapunov-Krasovskii functional
\begin{equation}\label{eq:LyapunovFunctional}
V(\bm{e_\tau}(t))=e(t)^T\!P\,e(t)\,+\int_{t-\tau}^{t}\!\!\!\!\!(\theta-t+\tau)\,\dot{e}(\theta)^T\!R\,\dot{e}(\theta)d\theta,
\end{equation}
with $P>0$, $R>0$ satisfying the LMI
\begin{equation}\label{eq:LMI}
\begin{bmatrix}
\Phi & P-\Psi_2^T+(A+BK)^T\Psi_3 & -\tau\Psi_2^TBK\\
* & -\Psi_3-\Psi_3^T+\tau R & -\tau\Psi_3^TBK\\
* & * & -\tau R
\end{bmatrix}<0,
\end{equation}
where $\Phi=(A+BK)^T\Psi_2+\Psi_2^T(A+BK)$, and $\Psi_2$, $\Psi_3$ are slack variables. Once again, it is worth noting that the existence of a Lyapunov-Krasovskii functional in the form \eqref{eq:LyapunovFunctional} is only a \emph{sufficient} condition for GAS. As a result, the literature on time-delayed systems presents several extensions aimed at reducing the restrictiveness of the stability conditions \eqref{eq:LMI}, see e.g. \cite{Wang2007}. Although this paper will not address other Lyapunov-Krasovskii functionals for the sake of simplicity, it is worth noting that the constraint enforcement strategy presented in the next section can be easily extended to other LTD stability results.

\section{Auxiliary Control}\label{sec:AuxiliaryControl}
The objective of this section is to augment the pre-stabilized system with an \emph{add-on} control loop that ensures constraint satisfaction. 
Motivated by the delay-free case derived in \cite{ERGbasic}, the Explicit Reference Governor is a constraint enforcement strategy that consists in assigning the dynamics of the auxiliary reference $v(t)$ as
\begin{equation}\label{eq:ERG}
\dot{v}=\Delta(\bm{x_\tau}(t),\bm{v_\tau}(t))\rho(v,r),
\end{equation}
where $\bm{x_\tau}(t)$, $\bm{v_\tau}(t)$ are analogous to \eqref{eq:Def_eTau}, and $\Delta(\bm{x_\tau},\bm{v_\tau})$, $\rho(v,r)$ satisfy the following definitions.
\smallskip
\begin{dfn}\label{def: Dynamic Safety Margin}
Let the prestabilized system \eqref{eq:PrestabilizedSystem} be subject to constraints \eqref{eq:OriginalConstraints}, let $\hat{v}(\theta\,|\,\bm{v_\tau(t)})$ be 
\begin{equation}\label{eq:PredictedReference}
\begin{cases}
\hat{v}(\theta)=v(t), &~\forall\theta\geq0,\\
\hat{v}(\theta)=v(t+\theta), &~\forall\theta\in[-\tau,0\,).
\end{cases}
\end{equation}
and let $\hat{x}(\theta\,|\,\bm{x_\tau(t)},\bm{v_\tau(t)})$ denote the solution to
\begin{equation}\label{eq:ForwardPredictions}
\begin{cases}
\dot{\hat{x}}(\theta)=A\hat{x}(\theta)+B\hat{u}(\theta-\tau),&~\forall\theta\geq0,\\
\hat{x}(\theta)=x(t+\theta), &~\forall\theta\in[-\tau,0\,),
\end{cases}
\end{equation}
where
\begin{equation}\label{eq:ForwardInput}
\hat{u}(\theta)=\bar{u}_{\hat{v}(\theta)}+K(\hat{x}(\theta)-\bar{u}_{\hat{v}(\theta)}),\qquad\forall\theta\geq-\tau.
\end{equation}
Given an auxiliary reference $v$ strictly satisfying constraints \eqref{eq:OriginalConstraints}, i.e.
\begin{equation}\label{eq:StrictContraints}
h_{i,x}^T\,\bar{x}_v+h_{i,u}^T\,\bar{u}_v+g_i>0,\quad i=1,\ldots,n_c,
\end{equation}
a continuous functional $\Delta:\real^n_{[-\tau,0]}\times\real^p_{[-\tau,0]}\to\real$ is a \textbf{``Dynamic Safety Margin''} if the following properties hold
\begin{subequations}\label{eq:DSM}
\begin{align}
&\Delta(\bm{x_\tau},\bm{v_\tau})=0~~\Rightarrow~~\Delta(\bm{\hat{x}_\tau(\theta)},\bm{\hat{v}_\tau(\theta)})\geq0,\label{eq:DSM_Invariant}\\
&\Delta(\bm{x_\tau},\bm{v_\tau})\geq0~~\Rightarrow~~ h_{x,i}^T\hat{x}(\theta)+h_{u,i}^T\hat{u}(\theta)+g_i\geq0,\label{eq:DSM_Positive}\\
&\Delta(\bm{x_\tau},\bm{v_\tau})>0~~\Rightarrow~~ h_{x,i}^T\hat{x}(\theta)+h_{u,i}^T\hat{u}(\theta)+g_i>0,\label{eq:DSM_StrictlyPositive}\\
&h_{x,i}^T\bar{x}_v+h_{u,i}^T\bar{u}_v+g_i\geq\delta~~\Rightarrow~~\Delta(\bar{x}_v,v)\geq\varepsilon,\!\!\label{eq:DSM_Returnable}
\end{align}
\end{subequations}
for all $\theta\geq0$, for all $i=1,\ldots,n_c$, and for a suitable value $\varepsilon>0$, given any $\delta>0$ hereafter denoted as a ``static safety margin''. \sqr
\end{dfn}
\smallskip
\begin{dfn}\label{def: Attraction Field}
Given the steady-state constraints \eqref{eq:StrictContraints} and a static safety margin $\delta>0$, a bounded and piecewise continuous function $\rho:\real^p\times\real^p\to\real^p$ is an \textbf{``Attraction Field''} if, for any initial condition $v(0)$ satisfying $h_{i,x}^T\,\bar{x}_{v(0)}+h_{i,u}^T\,\bar{u}_{v(0)}+g_i\geq\delta$, the system $\dot{\hat{v}}=\rho(r,\hat{v})$ is such that
\begin{enumerate}[1.]
\item For any piecewise continuous signal $r(t)\in\real^p$, $h_{i,x}^T\,\bar{x}_{v}+h_{i,u}^T\,\bar{u}_{v}+g_i\geq\delta$, $i=1,\ldots,n_c$;
\item Given a constant reference satisfying $h_{i,x}^T\,\bar{x}_{r}+h_{i,u}^T\,\bar{u}_{r}+g_i\geq\delta$, $i=1,\ldots,n_c$, $\hat{v}=r$ is an Asymptotically Stable equilibrium of $\dot{\hat{v}}=\rho(r,\hat{v})$.\sqr
\end{enumerate}
\end{dfn}
\smallskip
The idea behind the ERG formula \eqref{eq:ERG} is that, whenever $\Delta(\bm{x_\tau},\bm{v_\tau})=0$, it is possible to guarantee constraint satisfaction by assigning $\dot{v}=0$. This is due to the fact that \eqref{eq:PredictedReference}-\eqref{eq:ForwardPredictions} account for the future trajectories of the system if the current value of the reference $v$ were to remain constant. Due to continuity, it is therefore possible to assign $\dot{v}\neq0$ without running any risk of constraint violation as long as $\Delta(\bm{x_\tau},\bm{v_\tau})>0$. Finally, convergence to the desired reference $r$ is ensured by the fact that, for any strictly steady-state admissible reference $v$, the condition $\Delta(\bm{x_\tau},\bm{v_\tau})=0$ cannot hold indefinitely due to \eqref{eq:DSM_Returnable}, meaning that $v(t)$ will eventually converge to the equilibrium point of the vector field $\rho(v,r)$, which is $r$. Detailed proofs for the ideas behind these argumentations can be found in e.g. \cite{ERGrobust}. The following subsections will illustrate how to construct a suitable dynamic safety margin and attraction field for LTD systems subject to linear constraints.

\subsection{Dynamic Safety Margin}
A possible way to satisfy requirements \eqref{eq:DSM} would be to use the minimal distance between the predicted trajectory and the constraints, meaning
\begin{equation}\label{eq:DSM_FullTrajectory}
\Delta(\bm{x_\tau},\bm{v_\tau})= \min_{\begin{smallmatrix}
i=1,\ldots,l\\ \theta\geq0
\end{smallmatrix}}\{h_{x,i}^T\hat{x}(\theta)+h_{u,i}^T\hat{u}(\theta)+g_i\}.
\end{equation}
The main issue with equation \eqref{eq:DSM_FullTrajectory} is that the trajectories $\hat{x}(\theta)$, $\hat{u}(\theta)$ would have to be computed over the infinite horizon $\theta\geq0$. As detailed in the following proposition, this limitation can be overcome by computing the trajectory over a finite window of duration $T\geq\tau$ and bounding the remaining trajectories using Lyapunov level-sets.\smallskip
\begin{prp}\label{prp:DSM}
Consider the trajectories \eqref{eq:PredictedReference}-\eqref{eq:ForwardInput}, and let $T\geq\tau$ be a prediction horizon. Given 
\begin{equation}\label{eq:DSM_HorizonTrajectory}
\Delta_T(\bm{x_\tau},\bm{v_\tau})=\min_{\begin{smallmatrix}
i=1,\ldots,l\\ \theta\in[0,T]
\end{smallmatrix}}\{h_{x,i}^T\hat{x}(\theta)+h_{u,i}^T\hat{u}(\theta)+g_i\},
\end{equation}
and given
\begin{equation}\label{eq:DSM_Lyapunov}
\Delta_\infty(\bm{x_\tau},\bm{v_\tau})= \Gamma(v)-V(\bm{\hat{e}_\tau}(T)),
\end{equation}
where $V(\bm{\hat{e}_\tau}(T))$ is a Lyapunov-Krasovskii functional evaluated on $\bm{\hat{e}_\tau}(T)=\bm{\hat{x}_\tau}(T)-v$, and $\Gamma(v)$ is a threshold value satisfying
\begin{equation}\label{eq:ThresholdValue}
V(\bm{\hat{e}_\tau}(T)) \leq \Gamma(v) ~\Rightarrow~h_{x,i}^T\hat{x}(\theta)+h_{u,i}^T\hat{u}(\theta)+g_i,~\forall \theta>T,
\end{equation}
for $i=1,\ldots,n_c$, then
\begin{equation}\label{eq:DSM_Complete}
\Delta(\bm{x_\tau},\bm{v_\tau})= \min\{\kappa_1 \Delta_T(\bm{x_\tau},\bm{v_\tau})\:,\:\kappa_2\Delta_\infty(\bm{x_\tau},\bm{v_\tau})\},
\end{equation}
is a dynamic safety margin for any $\kappa_1>0$ and $\kappa_2>0$.\sqr
\end{prp}
\smallskip
\begin{proof}
For what concerns the time window $\theta\in[0,T]$, it is sufficient to note that \eqref{eq:DSM_HorizonTrajectory} captures the minimum distance between the trajectory of \eqref{eq:PredictedReference}-\eqref{eq:ForwardInput} and the boundary of constraints \eqref{eq:OriginalConstraints}. As for the time window $\theta>T$, it follows from \eqref{eq:ForwardPredictions} that $\hat{v}(\theta)=v$ for $\theta\in[T-\tau,\infty)$. This implies that, at the end of the prediction horizon $T\geq\tau$, the remaining behavior will be consistent with those of an LTD system subject to a constant reference. Following from the results presented in Section \ref{sec:PrimaryControl}, the Lyapunov-Krasovskii functional satisfies $V(\bm{\hat{e}_\tau}(\theta))<V(\bm{\hat{e}_\tau}(T))$, for all $\theta>T$. This is sufficient to ensure constraint satisfaction due to the definition of the threshold value \eqref{eq:ThresholdValue}.
\end{proof}
\smallskip
The basic idea behind Proposition \ref{prp:DSM} is that, although the auxiliary reference $v$ is a time-varying signal, its past history is only important in the time window $\theta\in[0,\tau]$. The properties discussed in Section \ref{sec:PrimaryControl} can thus be recovered by ``freezing'' the current reference and performing forward predictions to account for the past history of the reference. The remaining trajectories can then be bounded using the level-sets of Lyapunov-Krasovskii functionals. \smallskip

The final step for determining \eqref{eq:DSM_Complete} is to compute a suitable threshold value $\Gamma(v)$, which can be interpreted as the value of a level-set that is entirely contained in the constraints. To do so, it is sufficient to note that the presented Lyapunov-Krasovskii functionals \eqref{eq:LyapRazumikhin}, \eqref{eq:LyapFunctional_Q}, and \eqref{eq:LyapunovFunctional} all satisfy the quadratic lower bound 
\begin{equation}\label{eq:LyapLowerBound}
V(\bm{\hat{e}_\tau}(T))\geq \hat{e}(T)^T \!P \hat{e}(T).
\end{equation}
As a result, it follows from \cite{ERGlin} that \eqref{eq:ThresholdValue} can be satisfied for a given constraint \eqref{eq:OriginalConstraints} for
\begin{equation}\label{eq:Gamma}
\Gamma_i(v)=\frac{(h_{x,i}^T\bar{x}_v+h_{u,i}^T\bar{u}_v+g_i)^2}{(h_{x,i}^T+h_{u,i}^TK)P^{-1}(h_{x,i}+h_{u,i}K)}.
\end{equation}
As a consequence, considering the collection of all the constraints, it follows that $\Gamma(v)=\min\{\Gamma_i(v)\}$ for $i=1,\ldots,n_c$. Note that this approach is valid for any Lyapunov-Krasovskii functional that satisfies \eqref{eq:LyapLowerBound}. Since this is true for most results pertaining to LTD systems, e.g. all the formulations provided in \cite{Fridman2014}, the constraint handling strategy presented in this paper is easily generalizable to account for more advanced Lyapunov-Krasovskii functionals.
\smallskip

\begin{rmk}
It is worth noting that the matrix $P$ in equations \eqref{eq:LyapRazumikhin}, \eqref{eq:LyapFunctional_Q}, and \eqref{eq:LyapunovFunctional} can be chosen freely as long as the corresponding LMI is satisfied. In the case of Linear Time-Invariant systems, it has been shown in \cite{ERGlin} that, given a constraint $h_{x}(\bar{x}_v+e)+h_{u}(\bar{u}_v+Ke) + g\geq0$, it is possible to maximize the ERG performance by computing $P$ as the solution to
\begin{equation}\label{eq:OptimalInvSet}
\begin{cases}
\min&\!\!\!\log\det P\\
\text{s.t.}& \!\!\!P\geq (h_{x,i}+K^Th_{u,i})(h_{x,i}^T+h_{u,i}^TK),\\
&\!\!\!P>0,\\
& \!\!\!(A+BK)^TP+P(A+BK)\!<\!0.\\
\end{cases}
\end{equation}
This is justified by the idea that $P$ should be chosen so that the volume of the largest Lyapunov level-set compatible with the constraints is as aligned as possible to the constraint boundary while also satisfying \eqref{eq:LyapIneq}. Although in the case of time delay systems it is somewhat less clear how the choice of $P$ influences the performance of the ERG, a practical approach for selecting the matrix $P$ (as well as the auxiliary variables $q>0$, $Q>0$, $R>0$) is to solve the optimization problem \eqref{eq:OptimalInvSet}, where the last constraint is substituted with the LMI \eqref{eq:LMI_Razumikhin}, \eqref{eq:LMI_LKQ}, or \eqref{eq:LMI} depending on the selected Lyapunov-Krasovskii functional. \sqr
\end{rmk}

\subsection{Attraction Field}
This section addresses the design of the attraction field $\rho(v,r)$. Since this element of the ERG does not depend on the system delays, it is possible to use the standard attraction/repulsion approach \cite{PF}. Therefore, a possible solution is to use
\begin{equation}\label{eq: Attraction Field}
\rho(v,r)=\rho_0(v,r)+\sum_{i=1}^l\rho_i(v),
\end{equation}
with
\begin{equation}\label{eq: Attraction Term}
\rho_0(v,r)=\frac{r-v}{\min\{\|v-r\|\,,\,\eta\}},
\end{equation}
where $\eta>0$ is an attraction term that points towards $r\in\real^p$, and 
\begin{equation}\label{eq: Repulsion Term}
\rho_i(v)\!=\!\max\!\left\{\!\frac{\zeta-(h_{x,i}^T\bar{x}_v+h_{u,i}^T\bar{u}_v+g_i)}{\zeta-\delta},0\!\right\}\frac{h_{x,i}+K^Th_{u,i}}{\|h_{x,i}+K^Th_{u,i}\|},
\end{equation}
where $\zeta>\delta$ is a repulsion term that points inside the $i$th constraint.

\subsection{Main Result}
By combining all the previous components, the following proposition formally states the properties of the proposed constrained control architecture.\smallskip

\begin{prp}
Let \eqref{eq:System} be a time-delayed linear system subject to a constant delay $\tau$, and let \eqref{eq:OriginalConstraints} be a nonempty set of linear state and input constraints. Moreover, given a primary control law \eqref{eq:PrimaryControl}, let $V(\bm{e_\tau})$ be a Lyapunov-Krasovskii functional, and let \eqref{eq:DSM_Complete} be the associated dynamic safety margin. Then, given the attraction field \eqref{eq: Attraction Field}, and given an initial auxiliary reference $v(0)$ such that $\Delta(\bm{e_\tau}(0),v(0))\geq0$, the Explicit Reference Governor \eqref{eq:ERG} ensures the following properties:
\begin{enumerate}[1.]
\item For any piecewise continuous reference $r(t)\in\real^p$ Constraints \eqref{eq:OriginalConstraints} are satisfied for all time instants $t\geq0$;
\item For any constant reference $r\in\real^p$ that strictly satisfies \eqref{eq:OriginalConstraints}, the system output $y(t)$ asymptotically tends to $r$.\sqr
\end{enumerate}
\end{prp}
\smallskip
\begin{proof}
By construction, \eqref{eq:DSM_Complete} and \eqref{eq: Attraction Field} satisfy the requirements of Definitions \ref{def: Dynamic Safety Margin} and \ref{def: Attraction Field}, respectively. Therefore, the statement follows directly from the ERG framework detailed in \cite{ERGbasic}.
\end{proof}
\smallskip

It is worth noting that the systematic implementation of the ERG framework to time-delayed systems is only limited by the availability of suitable Lyapunov-Krasovskii functionals for the primary control loop. Indeed, the main interest in the scheme is that if the system can be stabilized without taking constraints into account, then the ERG can be used as an \textit{add-on} unit for handling the constraints. The following section illustrates how the results presented in this paper can be implemented in practice and also shows that the overall performances of the controlled systems will depend on the selected Lyapunov-Krasovskii functional.


\section{Experimental Validation}
\begin{figure}
\includegraphics[width=\columnwidth]{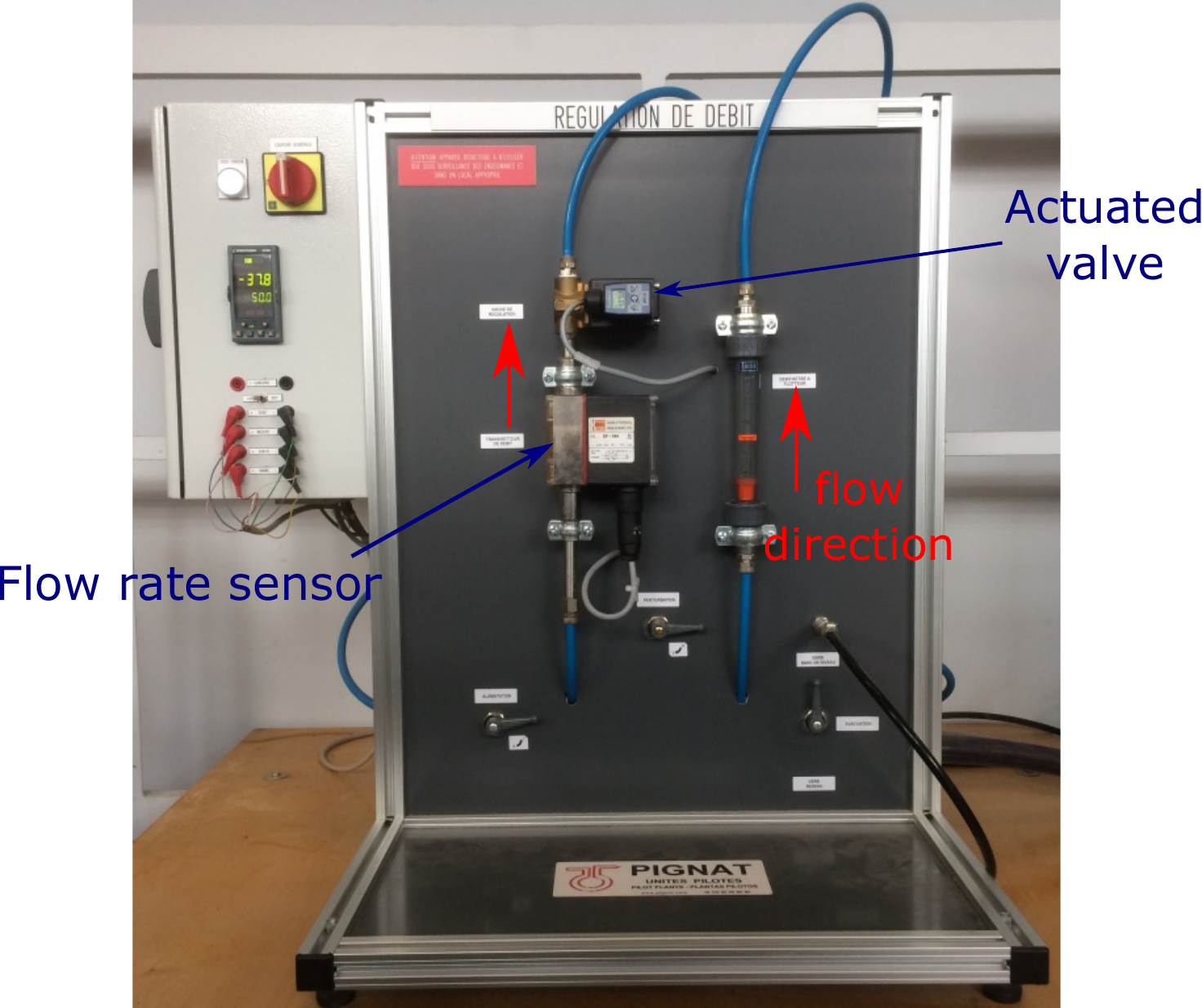}
\caption{Experimental setup.}\label{fig:testbed}
\end{figure}

To validate the results stated in this paper, the proposed strategy is used to control the water flow in a pipe through an actuated valve. The experimental setup is composed of a flow rate sensor Kobold type DF-MA and of an actuated valve Burkert type 8605 (see Fig. \ref{fig:testbed}).
The identified dynamics for the open-loop system are
\[
\dot{x}(t)=ax(t)+bu(t-\tau),
\]
where $x$ is the water flow rate [l/h], $u$ is the opening percentage of the valve [\%], $\tau=0.8\,[s]$ is the delay, $a=-0.82 \,[\text{s}^{-1}]$, and $b=0.7279\,[\text{lh}^{-1}\text{s}^{-1}]$. The system is required to go from the starting condition $x(0)=0\,[\text{l/h}]$ to the desired set-point $r=26\,[\text{l/h}]$ without violating the constraint $x\leq26.6\,[\text{l/h}]$.


The proposed primary control law is
\[
u(t)=\bar{u}_v+k(x(t)-v).
\]
The stability of the closed-loop system depends on the value of the control gain $k$. In particular, the following cases hold:
\begin{itemize}
\item For any $k\in[-1.12,1.12]$, the LMI \eqref{eq:LMI_Razumikhin} admits a solution, thus implying that \eqref{eq:LyapRazumikhin}-\eqref{eq:LyapFunctional_Q} are suitable Lyapunov functionals;
\item For any $k\in[-1.77,0)$, the LMI \eqref{eq:LMI_LKQ} admits a solution, thus implying that \eqref{eq:LyapunovFunctional} is a suitable Lyapunov functional;
\item For any $k\in(-3.54,-1.77)$, asymptotic stability may be proven using other Lyapunov functionals developed in the LTD literature, see e.g. \cite{Fridman2014};
\item For any $k<-3.54$ or $k>1.12$, the system is unstable.
\end{itemize}
To study the behavior of designs based on both the Delay-Independent and the Delay-Dependent conditions, the Explicit Reference Governor was tested for two different values of the control gain. In both cases, the experimental results were 
compared to the results of the numerical simulations.

\subsection{Mild Control Action}
\begin{figure}
\center
\includegraphics[width=\columnwidth]{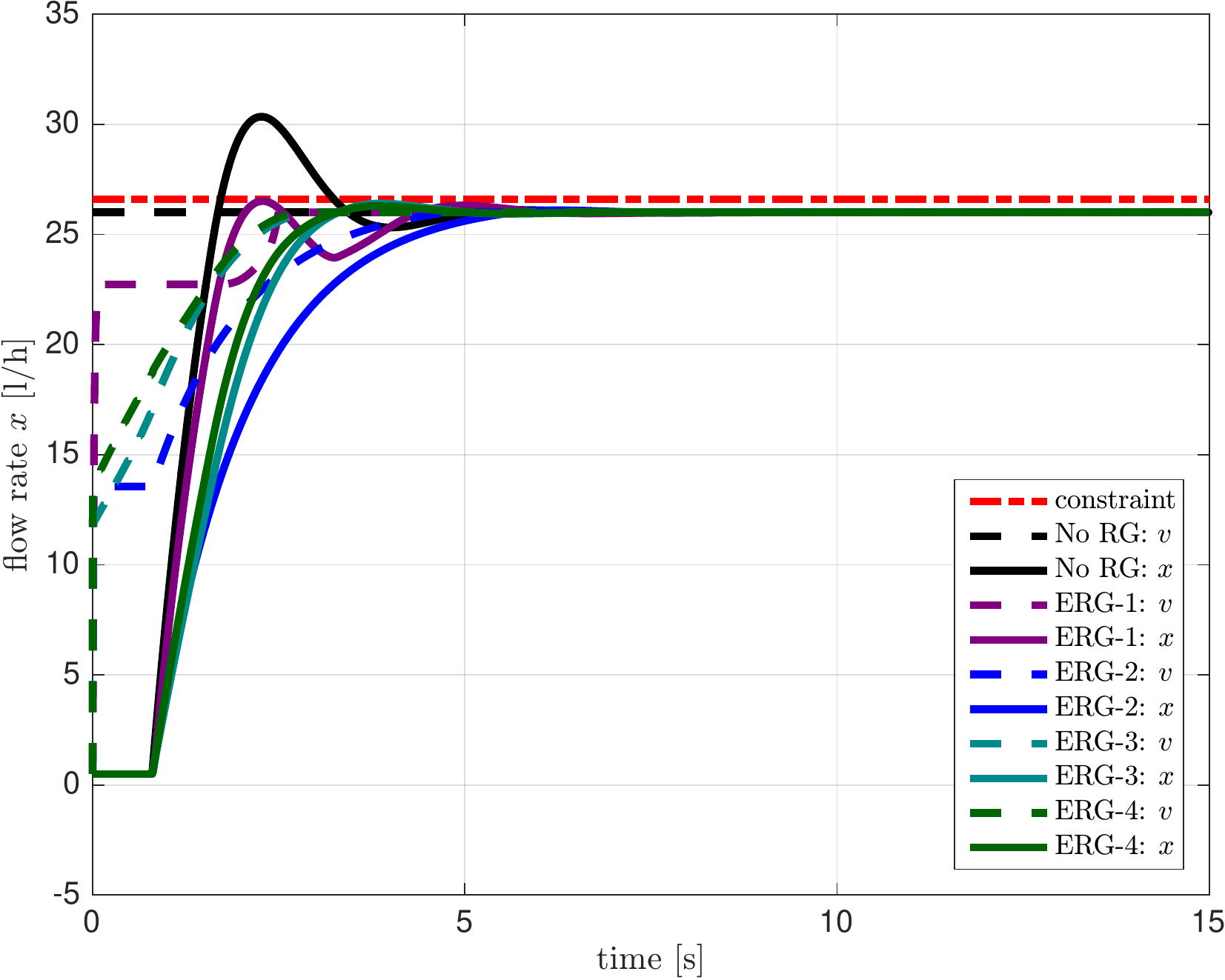}
\vspace{-0.5 cm}
\caption{Numerical simulations of the closed-loop response for the case $k=-1$. The dashed lines represent the auxiliary references $v(t)$, whereas the solid lines represent the state $x(t)$. The constraint boundary is represented by the dotted red line.\label{fig:Ex1_Numerical}}
\end{figure}
\begin{figure}
\center
\includegraphics[width=\columnwidth]{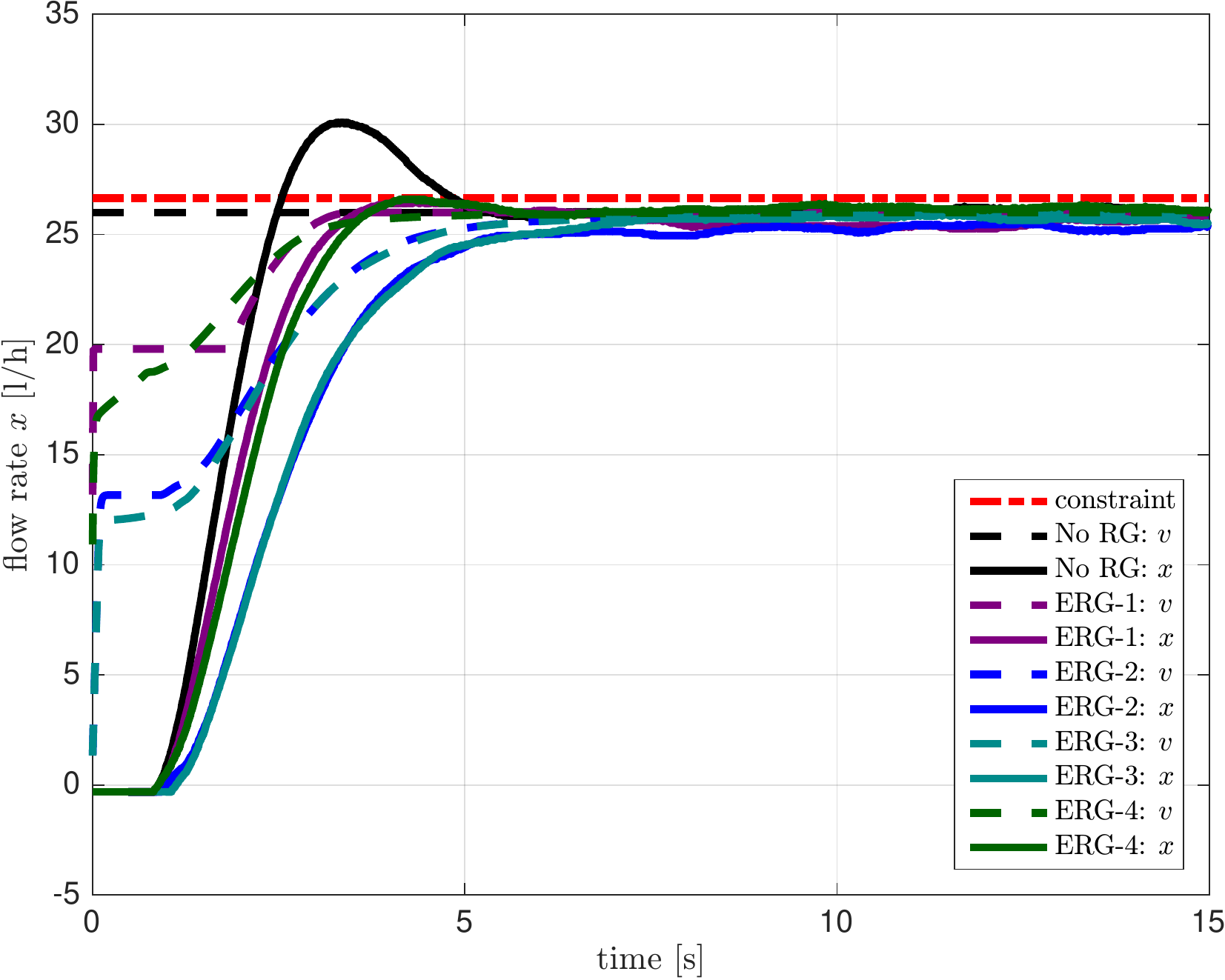}
\caption{Experimental results of the closed-loop response for the case $k=-1$. The dashed lines represent the auxiliary references $v(t)$, whereas the solid lines represent the state $x(t)$. The constraint boundary is represented by the dotted red line.\label{fig:Ex1_Experimental}}
\end{figure}

Given $k=-1$, the Lyapunov functionals \eqref{eq:LyapRazumikhin}-\eqref{eq:LyapFunctional_Q}, and \eqref{eq:LyapunovFunctional} can be constructed for $P=1$, $Q=0.86$, and $R=0.95$. Figures \ref{fig:Ex1_Numerical} and \ref{fig:Ex1_Experimental} provide the numerical and experimental comparisons, given
\begin{itemize}
\item \textbf{No Reference Governor (No RG):} The closed-loop system is directly subject to a step variation of the desired reference;
\item \textbf{``Infinite'' Horizon ERG (ERG-1):} The auxiliary reference is issued by an ERG, where \eqref{eq:DSM_Complete} is computed with $T=7s$ and $\kappa_1=50$. Due to the sizable length of the prediction horizon, the terminal dynamic safety margin $\Delta_\infty$ is omitted;
\item \textbf{Razumikhin-Based ERG (ERG-2):} The auxiliary reference is issued by an ERG, where \eqref{eq:DSM_Complete} is computed with $T=0.7s$, $\kappa_1=50$, $\kappa_2=20$. The terminal dynamic safety margin $\Delta_\infty$ is based on the Lyapunov-Razumikhin functional \eqref{eq:LyapRazumikhin};
\item \textbf{Krasovskii-Based ERG (ERG-3):} The auxiliary reference is issued by an ERG, where \eqref{eq:DSM_Complete} is computed with $T=0.7s$, $\kappa_1=50$, $\kappa_2=20$. The terminal dynamic safety margin $\Delta_\infty$ is based on the Lyapunov-Krasovskii functional \eqref{eq:LyapFunctional_Q};
\item \textbf{Krasovskii-Based ERG (ERG-4):} The auxiliary reference is issued by an ERG, where \eqref{eq:DSM_Complete} is computed with $T=0.7s$, $\kappa_1=50$, $\kappa_2=20$. The terminal dynamic safety margin $\Delta_\infty$ is based on the Lyapunov-Krasovskii functional \eqref{eq:LyapunovFunctional}.
\end{itemize}
As can be observed in the absence of a Reference governor, the closed-loop system violates the constraints due to the delay-induced overshoot. This issue is solved by augmenting the primary control law with any of the proposed Explicit Reference Governors.\smallskip

In terms of performance, the ``infinite'' horizon ERG achieves the fastest response time due to the fact that it takes into account the complete trajectory of the pre-stabilized system. Clearly, the main drawback of this method is that it requires the predictions over a long horizon, thus making it computationally expensive as well as increasingly susceptible to modeling errors.\smallskip

As for the other methods, the main advantage of performing shorter predictions and using a terminal set has the double advantage of reducing the computational footprint and increasing the overall robustness. Comparisons between the Razumikhin-based Lyapunov functional versus the Lyapunov-Krasovskii functionals show that the latter achieve a faster response. This is likely due to the fact that \eqref{eq:LyapRazumikhin} can remain constant up to the full time delay, whereas both functions \eqref{eq:LyapFunctional_Q} and \eqref{eq:LyapunovFunctional} are almost everywhere time-decreasing.

\subsection{Aggressive Control Action}
\begin{figure}
\center
\includegraphics[width=\columnwidth]{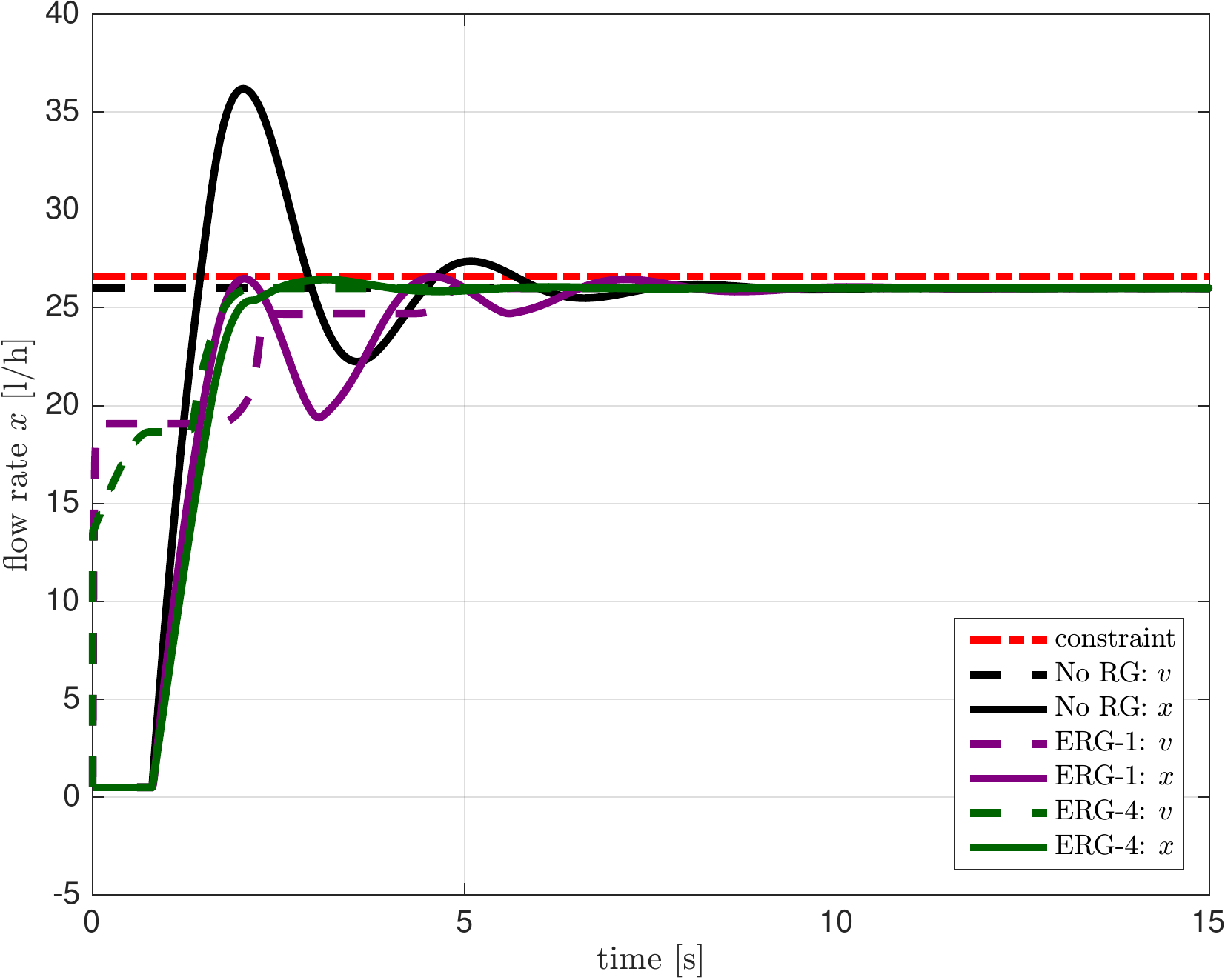}
\vspace{-0.5 cm}
\caption{Numerical simulations of the closed-loop response for the case $k=-1.68$. The dashed lines represent the auxiliary references $v(t)$, whereas the solid lines represent the state $x(t)$. The constraint boundary is represented by the dotted red line.\label{fig:Ex2_Numerical}}
\end{figure}

\begin{figure}
\center
\includegraphics[width=\columnwidth]{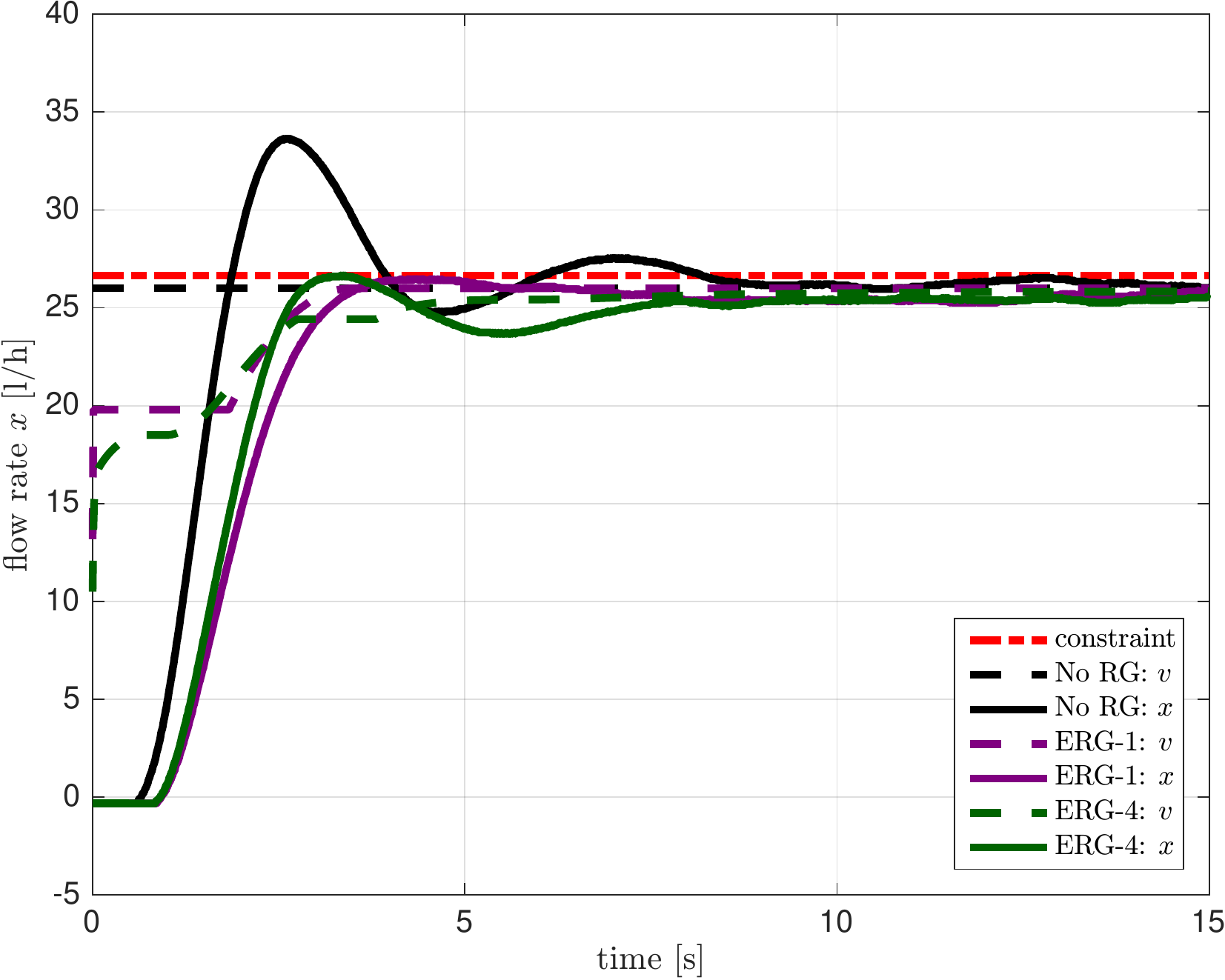}
\vspace{-0.5 cm}
\caption{Experimental results of the closed-loop response for the case $k=-1.68$. The dashed lines represent the auxiliary references $v(t)$, whereas the solid lines represent the state $x(t)$. The constraint boundary is represented by the dotted red line.\label{fig:Ex2_Experimental}}
\vspace{-0.5 cm}
\end{figure}

Given $k=-1.68$, it is not possible to prove asymptotic stability using delay-independent conditions. However, the Lyapunov functional \eqref{eq:LyapunovFunctional} holds for $P=1$ and $R=0.64$.
Figures \ref{fig:Ex2_Numerical} and \ref{fig:Ex2_Experimental} provide the numerical and experimental comparisons, given
\begin{itemize}
\item \textbf{No Reference Governor (No RG):} The closed-loop system is directly subject to a step variation of the desired reference;
\item \textbf{``Infinite'' Horizon ERG (ERG-1):} The auxiliary reference is issued by an ERG, where \eqref{eq:DSM_Complete} is computed with $T=7s$ and $\kappa_1=50$. Due to the sizable length of the prediction horizon, the terminal dynamic safety margin $\Delta_\infty$ is omitted;
\item \textbf{Krasovskii-Based ERG (ERG-4):} The auxiliary reference is issued by an ERG, where \eqref{eq:DSM_Complete} is computed with $T=0.7s$, $\kappa_1=50$, $\kappa_2=20$. The terminal dynamic safety margin $\Delta_\infty$ is based on the Lyapunov-Krasovskii functional \eqref{eq:LyapunovFunctional};
\end{itemize}

As in the previous case, the closed-loop system violates the constraints in the absence of a Reference governor. This issue is solved by augmenting the primary control law with an Explicit Reference Governor.\smallskip

As for the performance of the proposed ERG schemes, it is counter-intuitive to note that the response of the ``infinite'' horizon ERG is overall worse with respect to the Lyapunov-based ERG. This is likely due to the fact that, by moving quicker in the beginning, the ``infinite'' horizon approach causes a larger transient response and ends up having to issue a more conservative action later on.\smallskip

It is worth noting that, even though Figures \ref{fig:Ex1_Numerical}-\ref{fig:Ex1_Experimental} and Figures \ref{fig:Ex2_Numerical}-\ref{fig:Ex2_Experimental} present feature slightly different responses (likely due to model mismatch), constraint satisfaction is still guaranteed. As discussed in \cite{ERGrobust}, this is due to the inherent robustness of the Lyapunov-based ERG.



\section{Conclusions}
This paper proposed an explicit reference governor approach for the control of time delay systems subject to state and input constraints. The method consists in pre-stabilizing the system using a primary control loop and then introducing an auxiliary control loop that ensures constraint satisfaction by suitably manipulating the dynamics of the applied reference. The proposed scheme can be implemented using a variety of Lyapunov-Razumikhin and Lyapunov-Krasovskii functionals, although it has been shown in the numerical and experimental validations that the resulting performance will depend on the degree of conservativeness of the selected invariant.

\end{document}